%% file: ms.tex
\documentclass[journal]{IEEEtran}
\usepackage[caption=false,font=footnotesize]{subfig}
\usepackage[nopdftrailerid=true]{pdfprivacy}
\usepackage{booktabs}
\usepackage{datatool}
\usepackage{listings}
\usepackage{tikz}
\usetikzlibrary{positioning}
\usetikzlibrary{trees}
\usetikzlibrary{fit}
\usetikzlibrary{backgrounds}
\usepackage{hyperref}

\lstdefinestyle{lststyle}{basicstyle=\tt\footnotesize, captionpos=b, frame=single}

\begin{document}
\begin{filecontents*}{python/artifacts/keys-values.csv}
,key,value
0,epochs-num,20
1,batch-size,64
2,lr,0.01
\end{filecontents*}
\DTLloaddb{keys-values}{python/artifacts/keys-values.csv}

\title{A Makefile for Developing Containerized \LaTeX\ Technical Documents}

\author{Paschalis~Bizopoulos
\thanks{P. Bizopoulos is an Independent Researcher, Thessaloniki, Greece e-mail: pbizop@gmail.com}}

\maketitle

\begin{abstract}
	We propose a Makefile for developing containerized \LaTeX\ technical documents.
	The Makefile allows the author to execute the code that generates variables, tables and figures (results), which are then used during the \LaTeX\ compilation, to produce either the draft (fast) or full (slow) version of the document.
	We also present various utilities that aid in automating the results generation and improve the reproducibility of the document.
	We lastly release an open source repository of this paper that uses the Makefile.
\end{abstract}

\section{Introduction}
Developing technical documents that present both computationally generated results (variables, tables and figures) and natural text, is a central aspect of computational research.
One programming paradigm that was proposed as a solution is \textit{literate programming}~\cite{knuth1984literate}, in which the author inserts snippets of code and the output of its execution, alongside natural text.
Applications of \textit{literate programming} include Jupyter~\cite{kluyver2016jupyter} which allows publishing code, results and explanations in a executable format, PythonTeX~\cite{poore2015pythontex} which allows Python to be executed in \LaTeX\ and ActivePapers~\cite{hinsen2014activepapers} in which code can be executed using the Java Virtual Machine.

Although \textit{literate programming} provides many advantages for demonstrative projects and immediate visual feedback during experimentation, its `dual nature' constrains authors in using specifically designed text editors and/or source code formats thus leading to `app/vendor lock-in' situations.
Moreover, the overlap between natural text and results code makes \textit{literate programming} difficult to debug and version control.

We propose a Makefile for developing containerized \LaTeX\ technical documents.
By using portable technologies that withstood the test of time, we ensure the stability and portability of the proposed Makefile.
We write this paper using a repository that consists of the Makefile and choose Docker as the container technology and the Python programming language for generating the results.

For the rest of the paper we will refer to:
\begin{itemize}
	\item \textit{results code} as the scientific-oriented programming language code (\textit{main.py}) that produces the variables, tables and figures,
	\item \textit{results} as the variables, tables and figures that are generated by the \textit{results code},
	\item \textit{\LaTeX\ code} as the natural text (\textit{ms.tex}) and bibliography (\textit{ms.bib}),
	\item \textit{code} as both the \textit{results code} and \textit{\LaTeX\ code},
	\item \textit{document} as the technical document that contains the rendered \textit{results} and natural text and
	\item \textit{author} as the author of the \textit{document} that develops the \textit{code}.
\end{itemize}

\section{The Makefile}

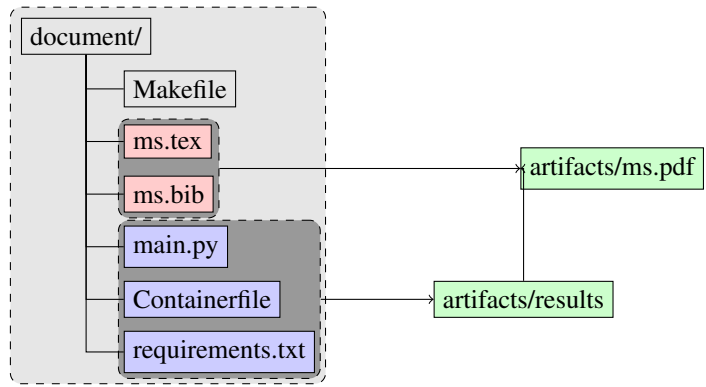
\begin{figure}[!t]
	\centering
	\begin{tikzpicture}[%
			grow via three points={one child at (0.5,-0.7) and two children at (0.5,-0.7) and (0.5,-1.4)},
			edge from parent path={(\tikzparentnode.south) |- (\tikzchildnode.west)}]
			\node[draw](document){document/}
			child{node[draw, anchor=west](makefile){Makefile}}
			child{node[draw, anchor=west, fill=red!20](tex){ms.tex}}
			child{node[draw, anchor=west, fill=red!20](bib){ms.bib}}
			child{node[draw, anchor=west, fill=blue!20](py){main.py}}
			child{node[draw, anchor=west, fill=blue!20](containerfile){Containerfile}}
			child{node[draw, anchor=west, fill=blue!20](requirements){requirements.txt}};
			\begin{scope}[on background layer]
				\node[draw, rounded corners, dashed, fill=gray!20, inner sep=4pt, fit=(document) (requirements)]{};
				\node[draw, rounded corners, dashed, fill=gray!80, inner sep=2pt, fit=(py) (requirements)](code){};
				\node[draw, rounded corners, dashed, fill=gray!80, inner sep=2pt, fit=(tex) (bib)](document){};
			\end{scope}
			\node[draw, right=1.5cm of code, fill=green!20] (results) {python/artifacts/results};
			\path[draw, ->] (code) -- (results);
			\node[draw, right=4cm of document, fill=green!20] (pdf) {python/artifacts/ms.pdf};
			\path[draw, ->] (results) |- (pdf);
			\path[draw, ->] (document) -- (pdf);
	\end{tikzpicture}
	\caption{The files of a repository that uses the Makefile is depicted in the light gray background and the arrows denote the containerized data flow towards generating the \textit{document}.
	Green indicates temporary cache or generated files, red \textit{\LaTeX\ code} files, blue \textit{results code} files and within the dark gray background, files that are developed by the \textit{author}.}\label{fig:repositoryfiles}
\end{figure}

The proposed Makefile can be used for developing \textit{documents} that consist of \textit{results} generated from \textit{results code} executed within a container.
The three tools that are used by the Makefile are the Make program which is a build automation tool, \LaTeX\ which is a typesetting system for publishing high quality research documents, and the container technology (such as Docker) which allows portable execution of \textit{code} between different hardware, operating systems and environments.

The pseudoalgorithm of the Makefile is as follows:
\begin{lstlisting}[language=bash, style=lststyle, caption={Pseudoalgorithm of the Makefile.}]
artifacts/$(document_pdf): $(latex_code) artifacts/$(results)
	run container latex

artifacts/$(results): $(results_code)
	build image $(results_image)
	run container $(results_image)

clean:
	rm -rf artifacts/
\end{lstlisting}

The default target runs the container for the \LaTeX\ compilation while the \textit{results} target builds the image and runs the container that generates the \textit{results}.
When the \textit{author} edits a file from \textit{results code} or \textit{\LaTeX\ code}, only the required compilation is automatically triggered with Make.
Forced compilations without change could be done using the \textit{touch} utility that updates the modification time of a file.

The basic commands of the Makefile are the following:
\begin{lstlisting}[language=bash, style=lststyle, caption={Basic Makefile commands.}]
make                # Generate draft artifacts (fast).\ 
make FULL=1         # Generate full artifacts (slow).\ 
make clean          # Remove artifacts/ directory.
\end{lstlisting}

The \textit{author} may use the Makefile for developing \textit{documents} in the following way:
\begin{enumerate}
	\item clones/downloads the repository which consists of the Makefile and the set of files as depicted in Fig.\ref{fig:repositoryfiles},
	\item creates and develops the \textit{results code}.
	\item creates and develops the \textit{\LaTeX\ code}.
	\item executes \textit{make} for fast development iterations and experimentation,
	\item executes \textit{make FULL=1} for distribution or publication and
	\item executes \textit{make clean} to restore to a clean state.
\end{enumerate}

The only requirements that the Makefile imposes on \textit{code} is the presence of \textit{results code} that generates the draft and full versions of the \textit{document} and the presence of \textit{results code} and \textit{\LaTeX\ code} for saving/loading the \textit{results} to/from \textit{artifacts/} respectively.
An example of a \textit{results code} that generates the draft version \textit{document} in the field of neural networks would be a conditional that sets the number of epochs and training/validation/test samples to a low value.

\section{The Makefile utilities}
In this section we present useful utilities that can be combined with the Makefile to automate \textit{results} generation and improve the reproducibility of the \textit{document}.

A useful \LaTeX\ package for automatically presenting variables from \textit{results} in \textit{\LaTeX\ code} is \textit{datatool} with the following example use:
\begin{lstlisting}[language=TeX, style=lststyle, caption={\LaTeX\ datatool example of loading a file that contains pairs of keys and values (artifacts/keys-values.csv) generated by a \textit{results code} and getting the value of a key named lr.}]
\DTLloaddb{keys-values}{artifacts/keys-values.csv}
\DTLfetch{keys-values}{key}{lr}{value}
\end{lstlisting}

When using Python, a useful function for \textit{results code} is \textit{pandas.DataFrame.to\_latex} which automatically converts a dataframe table to a \LaTeX\ table:
\begin{lstlisting}[language=python, style=lststyle, caption={Convert Pandas DataFrame (df) to \LaTeX\ table (artifacts/metrics.tex) in \textit{results code}.}]
df = pd.DataFrame(table)
df.to_latex("artifacts/metrics.tex", float_format="%.2f")
\end{lstlisting}

There are some considerations need to be taken into account when reproducibility is needed.
Regarding \LaTeX\, a culprit against reproducibility is the time-date metadata embedded in the pdf output.
These can be disabled using the following commands into the \textit{\LaTeX\ code} or as extra arguments during \LaTeX\ compilation (as done by default in the Makefile):
\begin{lstlisting}[language=TeX, style=lststyle, caption={\LaTeX\ pdf reproducibility commands.}]
\pdfinfoomitdate=1
\pdfsuppressptexinfo=-1
\pdftrailerid{}
\end{lstlisting}

Regarding the \textit{results code} for Python, we can achieve deterministic stochasticity by setting the random seeds to a specific value such as:
\begin{lstlisting}[language=python, style=lststyle, caption={Python reproducibility commands for popular libraries.}]
# build-in random module
random.seed(0)
# numpy
np.random.seed(0)
# tensorflow
tf.random.set_seed(0)
# pytorch
torch.backends.cudnn.benchmark = False
torch.backends.cudnn.deterministic = True
torch.cuda.manual_seed_all(0)
torch.manual_seed(0)
\end{lstlisting}

Additionally, an \textit{author} can use the \textit{cmp} utility to compare \textit{documents} (generated in a different session, operating system or underlying hardware) byte by byte, and diffoscope for identifying sources of non-determinism as follows:
\begin{lstlisting}[language=bash, style=lststyle, caption={Test draft document version reproducibility. This can also be used as a test script when pushing or pull requesting to a remote repository.}]
# Generate the first draft document version.\ 
make
# Backup the first resulting pdf.\ 
cp artifacts/ms.pdf artifacts/ms-previous.pdf
# Update the modification timestamp of main.py.\ 
touch main.py
# Generate the second draft document version.\ 
make
# Compare draft document versions byte by byte.\ 
cmp artifacts/ms.pdf artifacts/ms-previous.pdf
# Compare draft document versions
# in a human readable form.\ 
diffoscope artifacts/ms.pdf artifacts/ms-previous.pdf
\end{lstlisting}

\section{Example use case of the Makefile}
This section provides a use case of the Makefile and also serves as a manual for developing \LaTeX\ \textit{documents} with Python as the \textit{results code} programming language.
We train, validate and test MobilenetV2 neural networks~\cite{sandler2018mobilenetv2} in PyTorch, on MNIST~\cite{lecun2010mnist}, FashionMNIST~\cite{xiao2017fashion}, KMNIST~\cite{clanuwat2018deep} and QMNIST~\cite{yadav2019cold}.
We use the default train, validation and test datasets and compare the use of ReLU, ReLU6\cite{dahl2013improving} and SiLU\cite{elfwing2018sigmoid} activation functions in the MobilenetV2 model.

\begin{figure}[!t]
	\subfloat{\includegraphics[width=0.5\columnwidth]{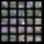}}
	\subfloat{\includegraphics[width=0.5\columnwidth]{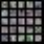}}
	\\
	\subfloat{\includegraphics[width=0.5\columnwidth]{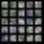}}
	\subfloat{\includegraphics[width=0.5\columnwidth]{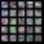}}
	\caption{The first 25 kernels of the first convolutional layer of MobilenetV2 with ReLU for MNIST (upper-left), FashionMNIST (upper-right), KMNIST (lower-left) and QMNIST (lower-right).}\label{fig:kernels}
\end{figure}

Making use of the \textit{datatool} package the values of the following variables are not directly referred in the main \textit{.tex} file but they are instead read by an intermediate \textit{.tex} file created by the \textit{results code}: the number of epochs is $\DTLfetch{keys-values}{key}{epochs-num}{value}$, the batch size is $\DTLfetch{keys-values}{key}{batch-size}{value}$ and the learning rate for SGD is $\DTLfetch{keys-values}{key}{lr}{value}$.
Additionally, making use of random seed setting we consistenly get the same figures as shown in Fig.\ref{fig:loss} and Fig.\ref{fig:kernels}.
We also use the \textit{pandas.DataFrame.to\_latex} command which automatically converts a dataframe table to a \LaTeX\ table (as shown in Table~\ref{table:table}):

\begin{figure}[!t]
	\subfloat{\includegraphics[width=0.5\columnwidth]{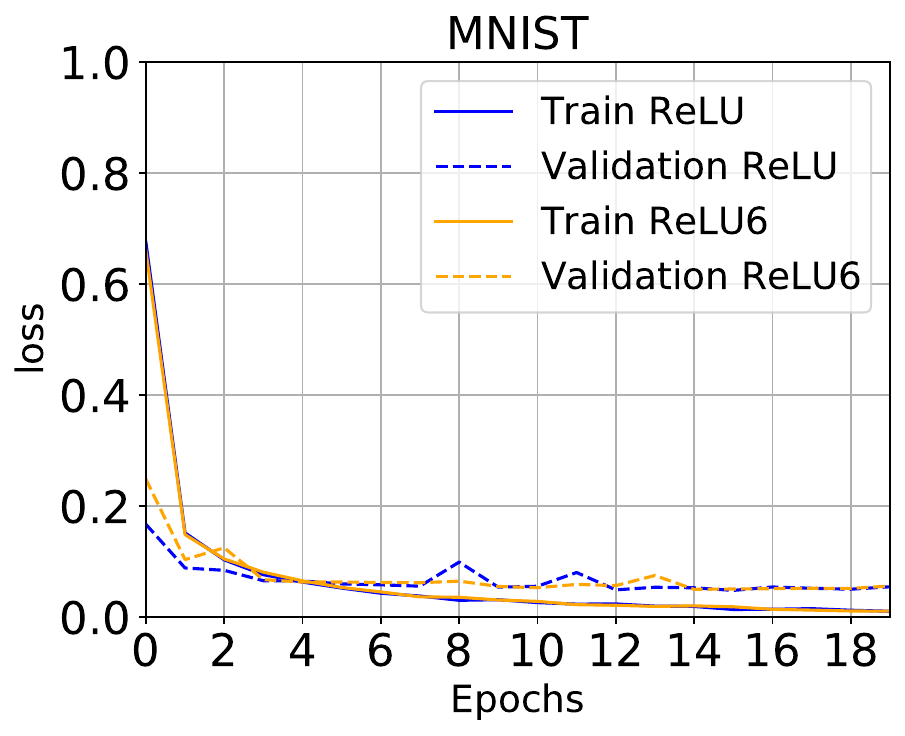}}
	\subfloat{\includegraphics[width=0.5\columnwidth]{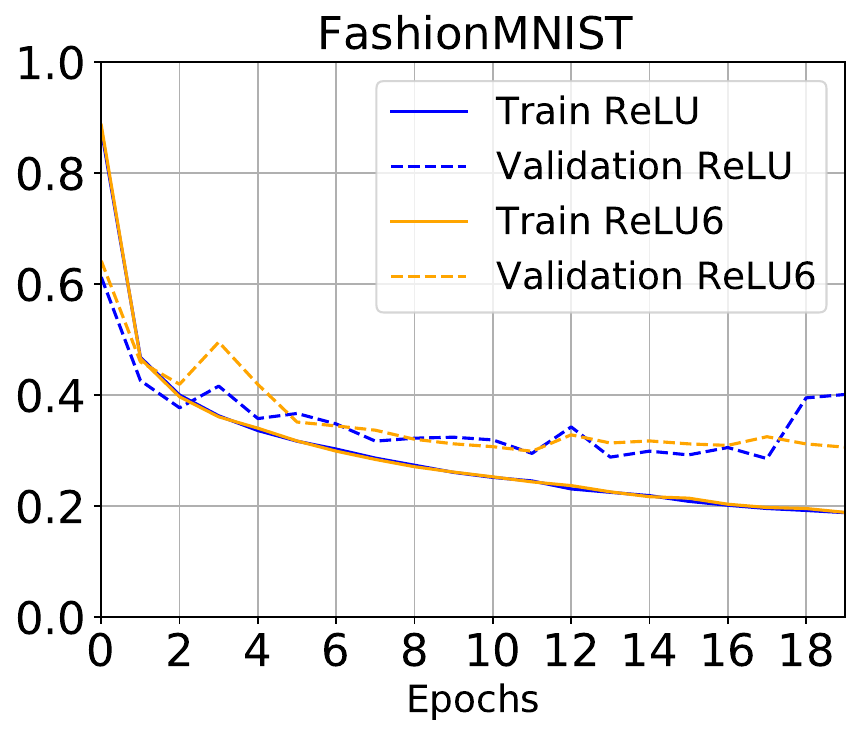}}
	\\
	\subfloat{\includegraphics[width=0.5\columnwidth]{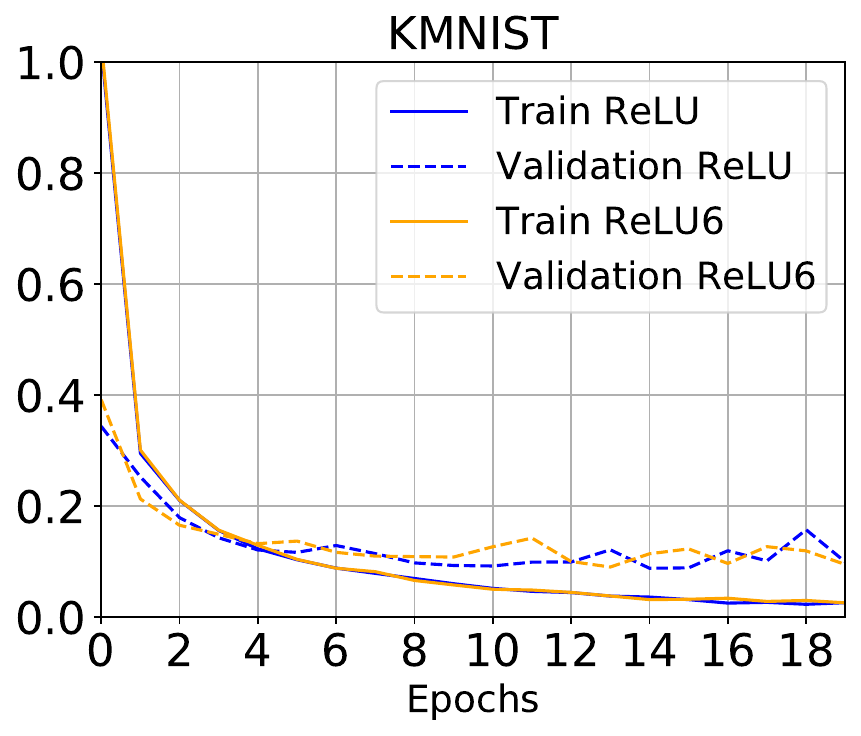}}
	\subfloat{\includegraphics[width=0.5\columnwidth]{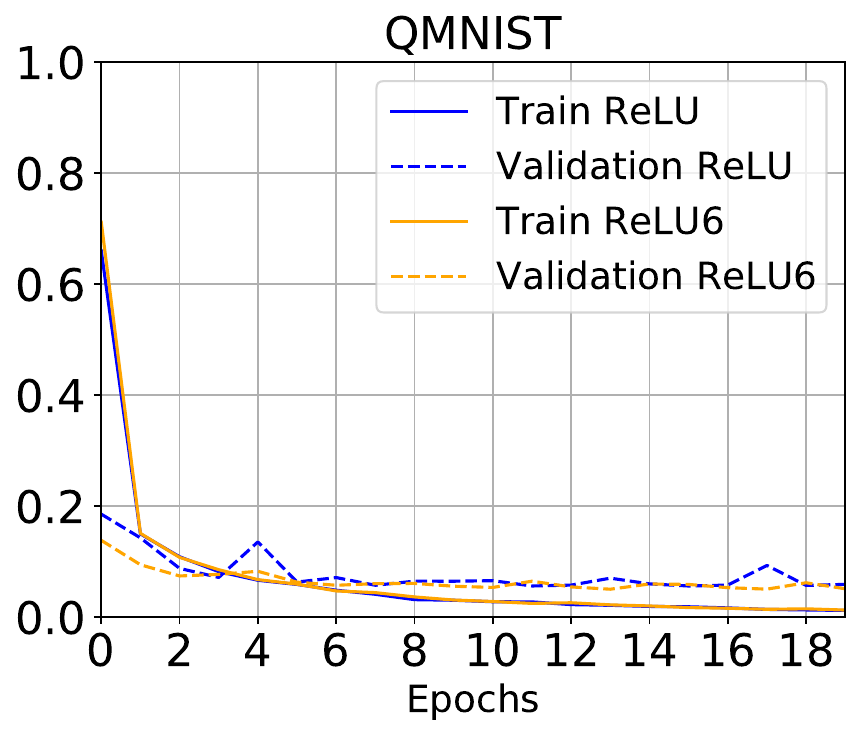}}
	\caption{Comparison of the train and validation loss of MobilenetV2 with ReLU, ReLU6 and SiLU for all datasets.}\label{fig:loss}
\end{figure}

\begin{table}[ht]
	\centering
	\caption{MobilenetV2 test dataset accuracies.}\label{table:table}
	\setlength\tabcolsep{4pt}
	\input{python/artifacts/metrics.tex}
\end{table}

\section{Discussion}
The proposed Makefile is based on plain text file editing and can be used with any operating system that supports containers and Make.
Make has existed for decades and has passed the test of time, while the container technology was standardized with the Open Container Initiative and there exist alternatives such as Podman that could be used as drop-in replacements for Docker.
Moreover the Makefile can be combined with any version control system, container registry provider and text editor thus preventing `app/vendor lock-in' situations.

Use cases of the Makefile include:
\begin{itemize}
	\item regression testing and debugging, to ensure that changes to \textit{code} do not alter \textit{results},
	\item common development environment across multiple \textit{authors},
	\item coauthors, reviewers, journal editors or other researchers can easily reproduce the \textit{document} with few requirements.
\end{itemize}

\section*{Conclusion}
Creation, usage and publication of technical documents are one of the top challenges for reproducible technical documents~\cite{barba2019praxis}.
The proposed Makefile aids to this purpose by providing a way to write reproducible technical documents using the standard tools that many of the academics already use (\LaTeX, Docker, Make) in a portable, efficient and future-proof way.

\bibliographystyle{IEEEtran}
\bibliography{ms.bib}

\end{document}

%% file: python/artifacts/metrics.tex
\begin{tabular}{lrrrr}
\toprule
{} &      MNIST & FashionMNIST &     KMNIST &     QMNIST \\
\midrule
\textbf{ReLU } &          98.74 &   \textbf{88.92} &          93.22 &          98.62 \\
\textbf{ReLU6} &          98.69 &            88.45 &          92.89 &          98.58 \\
\textbf{SiLU } & \textbf{98.85} &            88.08 & \textbf{93.59} & \textbf{98.63} \\
\bottomrule
\end{tabular}